\documentclass[aps,prb,reprint,preprintnumbers]{revtex4-2}
\usepackage[T1]{fontenc}
\usepackage{times}
\usepackage{graphicx,color}
\usepackage{amsfonts,amsmath,amssymb,amsbsy}
\usepackage[colorlinks=true, linkcolor=blue, citecolor=blue, urlcolor=blue]{hyperref}
\def\blue#1{\textcolor{blue}{#1}}
\begin{document}

\title{Structural transition of skyrmion quasiparticles under compression}

\author{Xichao Zhang}
\thanks{These authors contributed equally to this work.}
\affiliation{Department of Electrical and Computer Engineering, Shinshu University, 4-17-1 Wakasato, Nagano 380-8553, Japan}

\author{Jing Xia}
\thanks{These authors contributed equally to this work.}
\affiliation{Department of Electrical and Computer Engineering, Shinshu University, 4-17-1 Wakasato, Nagano 380-8553, Japan}

\author{Xiaoxi Liu}
\email[]{liu@cs.shinshu-u.ac.jp}
\affiliation{Department of Electrical and Computer Engineering, Shinshu University, 4-17-1 Wakasato, Nagano 380-8553, Japan}

\begin{abstract}
The competition between external drives and interparticle interactions are important for the static and dynamic properties of topological quasiparticles in confined geometries. Here, we study the dynamics of particlelike skyrmions under a current-induced compression, where skyrmions are stabilized in a confined space and form a skyrmion solid. We find that a lattice structural transition of nanoscale skyrmions could be induced by the compression, which depends on the initial lattice configuration and the compression level. The compression effect in the skyrmion solid with strong skyrmion-skyrmion repulsion could produce a wave front showing an abrupt change of the skyrmion density, which may propagate through the skyrmion solid depending on the compression level. Besides, a chain of skyrmions on the surface of the compressed skyrmion solid could show a fast and stable motion along the surface as long as it remains within the driving region. Some skyrmions may collapse during the compression due to the destructive skyrmion-skyrmion collision, which mainly happens in the area around the skyrmion injection boundary between the driving and compressive regions. Our results are useful for understanding the compression physics of quasiparticles carrying nontrivial topology.
\end{abstract}

\date{May 3, 2022}

\preprint{\href{https://doi.org/10.1103/PhysRevB.105.184402}{\textsl{Phys. Rev. B \textbf{105}, 184402 (2022)}}}

\maketitle

\section{Introduction}
\label{se:Introduction}

Nanoscale skyrmions in magnetic systems can be treated as quasiparticles~\cite{Bogdanov_1989,Roszler_NATURE2006,Lin_PRB2013,Reichhardt_PRL2015,Reichhardt_PRB2015A,Reichhardt_PRB2015B,Reichhardt_PRB2016,Reichhardt_NJP2016,Reichhardt_PRB2018,Reichhardt_PRL2018,Reichhardt_PRB2020,Vizarim_PRB2020,Reichhardt_2021}, which are promising information carriers that allow for a wide range of applications~\cite{Nagaosa_NNANO2013,Mochizuki_Review,Wiesendanger_Review2016,Finocchio_JPD2016,Kang_PIEEE2016,Kanazawa_AM2017,Wanjun_PHYSREP2017,Fert_NATREVMAT2017,Everschor_JAP2018,Bogdanov_NRP2020,Zhang_JPCM2020,Fujishiro_2020,Back_JPD2020,Gobel_PP2021,Li_MH2021,Luo_APLM2021,Tokura_CR2021,Vakili_JAP2021,Yu_JMMM2021,Marrows_APL2021,Li_SB2022}.
They are usually very rigid objects and can interact with each other effectively through the short-range skyrmion-skyrmion interaction and long-range dipolar interaction~\cite{Zhang_SR2015,Rozsa_PRL2016,Du_PRL2018,Brearton_PRB2020,Capic_JPCM2020,Lin_PRB2013}.
As the skyrmion carries a nonzero integer topological charge, its dynamics in a flat landscape are subject to both external driving forces and its topological nature~\cite{Nagaosa_NNANO2013,Mochizuki_Review,Wiesendanger_Review2016,Finocchio_JPD2016,Kang_PIEEE2016,Kanazawa_AM2017,Wanjun_PHYSREP2017,Fert_NATREVMAT2017,Everschor_JAP2018,Bogdanov_NRP2020,Zhang_JPCM2020,Fujishiro_2020,Back_JPD2020,Gobel_PP2021,Li_MH2021,Luo_APLM2021,Tokura_CR2021,Vakili_JAP2021,Yu_JMMM2021,Marrows_APL2021}.
Such a feature may lead to some unique dynamics compared to common particles. For example, a particlelike skyrmion driven by the spin current may move at an angle with respect to the driving force direction, which is called the skyrmion Hall effect~\cite{Zang_PRL2011,Wanjun_NPHYS2017,Litzius_NPHYS2017,Tang_NN2021,Tan_NC2021,Peng_NC2021,Reichhardt_NJP2016}.
The skyrmion Hall effect originates from the topological Magnus force acting on the skyrmion, of which the direction is perpendicular to the skyrmion velocity, and the sign depends on the topological charge~\cite{Zang_PRL2011,Wanjun_NPHYS2017,Litzius_NPHYS2017,Tang_NN2021,Tan_NC2021,Peng_NC2021,Reichhardt_NJP2016}.
Therefore, a system containing massive interacting skyrmions driven by an external force may show very different dynamic properties compared to that of other particles or quasiparticles.

Many skyrmions in a magnetic system with Dzyaloshinskii-Moriya (DM) interactions~\cite{Dzyaloshinsky,Moriya} may form a skyrmion solid, a skyrmion liquid, or a skyrmion gas, which depends on the density of skyrmions and their mutual interactions~\cite{Muhlbauer_SCIENCE2009,Yu_Nature2010,Zhou_NP2020,Huang_NN2020,Pinna_PRA2018}.
For skyrmions stabilized by interfacial and bulk DM interactions, the skyrmion-skyrmion interaction is usually repulsive, of which the magnitude is inversely proportional to the spacing between two skyrmions~\cite{Lin_PRB2013}.
Hence, such a repulsive skyrmion-skyrmion interaction plays an important role in the system of a skyrmion solid. It may affect the static lattice structure as well as the lattice dynamics of skyrmions. The static lattice structure of a skyrmion solid may be manipulated by an external force or field acting on all skyrmions~\cite{Karube_NM2016,Nakajima_SA2017,Walsem_PRB2019,Takagi_NC2020,Karube_PRB2020,Huang_NN2020}.
For example, Karube \textit{et al.} have demonstrated that the transition between a triangular skyrmion lattice and a square skyrmion lattice by varying the temperature and magnetic field~\cite{Karube_NM2016}.
However, it is unclear what will happen to the static lattice structure and lattice dynamics of a skyrmion solid when an external drive is only partially applied on a few skyrmions, while most other skyrmions serve as pinning obstacles due to the repulsive skyrmion-skyrmion interaction.

A possible route to study this problem is to compress a number of particlelike skyrmions using one boundary, which is inspired by recent reports on the compressional shocks in Yukawa solids~\cite{Feng_2019,Feng_2021A,Feng_2021B} and the report on the skyrmion avalanches~\cite{Reichhardt_PRL2018}.
In a two-dimensional (2D) Yukawa solid, the interaction between neighboring particles is described by the Yukawa potential~\cite{Yukawa_1947,Feng_2019,Feng_2021A,Feng_2021B}, which produces a strong attractive force between two adjacent particles when the interparticle distance is significantly small.
This is similar to a system of a massive number of nanoscale skyrmions stabilized by DM interactions, however, two rigid skyrmions experience a strong repulsive force when they are close to each other, which is described by the first-order modified Bessel function of the second kind~\cite{Lin_PRB2013}. 

In this work, we report the dynamics of particlelike skyrmions with an initial triangular or square lattice structure under compression.
We focus on the structural transition under compression, which is an important phenomenon of skyrmions~\cite{McDermott_2016A,McDermott_2016b,Karube_NM2016,Nakajima_SA2017,Walsem_PRB2019,Takagi_NC2020,Karube_PRB2020,Huang_NN2020}.
We find that by triggering the compression of skyrmion quasiparticles in a confined geometry through one boundary, one could modify the lattice structure of skyrmion quasiparticles in the compressive region.
We also find that a chain of fast moving skyrmions on the surface of the compressed skyrmion quasiparticles during the compression.

\begin{figure*}[t]
\centerline{\includegraphics[width=1.00\textwidth]{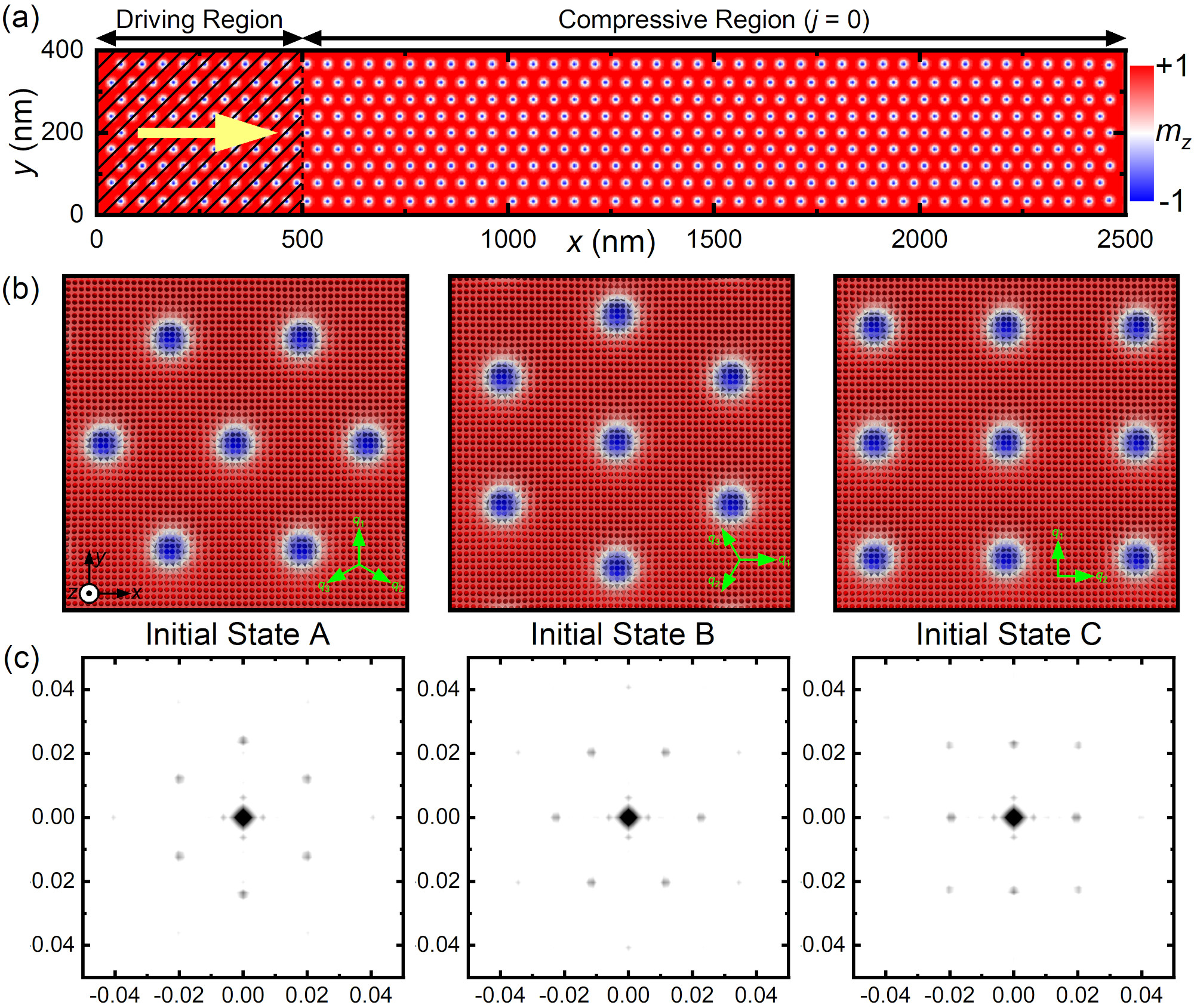}}
\caption{%
Simulation model and initial spin configurations.
(a) Top view of the monolayer nanotrack with a triangular lattice of nanoscale skyrmions as the initial spin configuration. The length, width, and thickness of the ferromagnetic monolayer equal $2500$ nm, $400$ nm, and $1$ nm, respectively. The dampinglike spin torque is only applied to the driving region between $x=0$ nm and $x=500$ nm. The region between $x=500$ nm and $x=2500$ nm is referred to as the compressive region, which is the focused region in this work. The color scale represents the out-of-plane spin component $m_z$, which has been used throughout the work. The yellow arrow indicates the general motion direction of skyrmions in the driving region upon the application of a driving force. 
(b) Close-up top views of three different initial spin configurations. Each cone represents a spin. The \textit{initial states A} and \textit{B} are two different triangular skyrmion lattices. The \textit{initial state C} is a square skyrmion lattice. The corresponding magnetic modulation vectors are given; see green arrows.
(c) The 2D Fourier transforms of the above three different initial spin configurations.
}
\label{FIG1}
\end{figure*}

\section{Methods}
\label{se:Methods}

We simulate the skyrmion system using the \textsc{mumax$^3$} micromagnetic simulator on a commercial graphics processing unit~\cite{MuMax}.
The spin dynamics of the system is governed by the Landau-Lifshitz-Gilbert (LLG) equation augmented with the dampinglike spin-orbit torque~\cite{Sampaio_NN2013,Tomasello_SREP2014,Wanjun_SCIENCE2015,Xichao_NC2016,Xichao_PRB2016A,Xichao_PRB2016B,Xia_APL2020},
\begin{equation}
\label{eq:LLGS-CPP}
\partial_{t}\boldsymbol{m}=-\gamma_{0}\boldsymbol{m}\times\boldsymbol{h}_{\text{eff}}+\alpha(\boldsymbol{m}\times\partial_{t}\boldsymbol{m})+\boldsymbol{\tau}_{\text{d}},
\end{equation}
where $\boldsymbol{m}$ is the normalized magnetization,
$t$ is the time,
$\gamma_0$ is the absolute gyromagnetic ratio,
$\alpha$ is the Gilbert damping parameter,
$\boldsymbol{h}_{\rm{eff}}=-\frac{1}{\mu_{0}M_{\text{S}}}\cdot\frac{\delta\varepsilon}{\delta\boldsymbol{m}}$ is the effective field with $\mu_{0}$, $M_{\text{S}}$, and $\varepsilon$ being the vacuum permeability constant, saturation magnetization, and average energy density, respectively.
$\boldsymbol{\tau}_{\text{d}}=u\left(\boldsymbol{m}\times\boldsymbol{p}\times\boldsymbol{m}\right)$ is the the dampinglike spin-orbit torque with the coefficient $u=\left|\left(\gamma_{0}\hbar/\mu_{0}e\right)\right|\cdot\left(j\theta_{\text{SH}}/2aM_{\text{S}}\right)$, where
$\hbar$ is the reduced Planck constant, $e$ is the electron charge, $a$ is the sample thickness, $j$ is the current density, and $\theta_{\text{SH}}$ is the spin Hall angle.
We ignore the fieldlike spin-orbit torque considering the fact that it does not affect the dynamics of rigid nanoscale skyrmion quasiparticles at a small driving force~\cite{Litzius_NPHYS2017}.
The micromagnetic energy terms included in our model are the ferromagnetic exchange energy, perpendicular magnetic anisotropy (PMA) energy, demagnetization energy, and interface-induced DM energy~\cite{Sampaio_NN2013,Tomasello_SREP2014,Xichao_NC2016,Xichao_PRB2016A,Xichao_PRB2016B,Xia_APL2020}.
The magnetic parameters are adopted from Refs.~\onlinecite{Sampaio_NN2013,Tomasello_SREP2014,Xichao_NC2016,Xichao_PRB2016A,Xichao_PRB2016B,Xia_APL2020}: $\gamma_{0}=2.211\times 10^{5}$ m A$^{-1}$ s$^{-1}$, $\alpha=0.3$, $M_{\text{S}}=580$ kA m$^{-1}$, exchange constant $A=15$ pJ m$^{-1}$, PMA constant $K=0.8$ MJ m$^{-3}$, and DM interaction constant $D=3$ mJ m$^{-2}$.
For simplicity, we assume that $\theta_{\text{SH}}=1$.
The mesh size is $2.5$ $\times$ $2.5$ $\times$ $1$ nm$^3$, which ensures both good computational accuracy and efficiency.

As depicted in Fig.~\ref{FIG1}(a), the sample is a nanotrack with a width of $400$ nm in the $y$ direction and a length of $2500$ nm in the $x$ direction. The nanotrack thickness equals $1$ nm.
We apply the periodic boundary conditions in the $\pm y$ directions, while open boundary conditions are applied in the $\pm x$ directions.
To compress the skyrmions, we assume that a straight current flows only in the underneath heavy metal on the left side of the sample ($x=0-500$ nm) and generates a uniform dampinglike spin-orbit torque on the above ferromagnetic region ranging between $x=0$ nm and $x=500$ nm, which could be realized in a ferromagnet/heavy metal device structure via the spin Hall effect~\cite{Sampaio_NN2013,Tomasello_SREP2014,Wanjun_SCIENCE2015,Xichao_NC2016,Xichao_PRB2016A,Xichao_PRB2016B,Xia_APL2020}.
With such a simplified injection geometry, the spin polarization orientation is assumed to be $\boldsymbol{p}=+\hat{x}$.

We consider three different initial spin configurations, which are illustrated in Fig.~\ref{FIG1}(b).
In the first spin configuration (i.e., the \textit{initial state A}), $441$ skyrmions are placed in the track and form a triangular lattice; also see Fig.~\ref{FIG1}(a).
The triangular skyrmion lattice can be described by the superposition of three magnetic modulation vectors $\textbf{\textit{q}}_{i}$ ($i=1,2,3$) and a uniform out-of-plane spin component~\cite{Karube_NM2016,Nakajima_SA2017,Walsem_PRB2019,Takagi_NC2020,Karube_PRB2020}; see green arrows in Fig.~\ref{FIG1}(b).
The second spin configuration (i.e., the \textit{initial state B}) is a triangular lattice of $456$ skyrmions, of which the orientation is rotated by $90$ degrees compared to the \textit{initial state A}.
The third spin configuration (i.e., the \textit{initial state C}) is a square lattice of $441$ skyrmions.
The square skyrmion lattice can be described by the superposition of two magnetic modulation vectors $\textbf{\textit{q}}_{i}$ ($i=1,2$) and a uniform out-of-plane spin component~\cite{Takagi_NC2020,Nakajima_SA2017}.
The 2D Fourier transforms of the above three different initial states are given in Fig.~\ref{FIG1}(c).

All these initial spin configurations are relaxed before the application of compression. Note that initially the system is in equilibrium and does not show interskyrmion compression as we set an appropriate spacing between adjacent skyrmions.
It should also be noted that the spacing between adjacent skyrmions at the relaxed initial state should not be much greater than the minimum required spacing, otherwise the compression effect may be insignificant.
For simplicity, we only study the system where all skyrmions have a topological charge of $Q=-1$ with $Q$ being defined as 
$Q=\frac{1}{4\pi}\int\boldsymbol{m}\cdot(\frac{\partial\boldsymbol{m}}{\partial x}\times\frac{\partial\boldsymbol{m}}{\partial y})dxdy$~\cite{Nagaosa_NNANO2013,Zhang_JPCM2020,Gobel_PP2021}.
Note that the total topological charge is calculated in a lattice-based approach by a built-in extension of \textsc{mumax$^3$}~\cite{MuMax,LatticeQ}.

\begin{figure}[t]
\centerline{\includegraphics[width=0.50\textwidth]{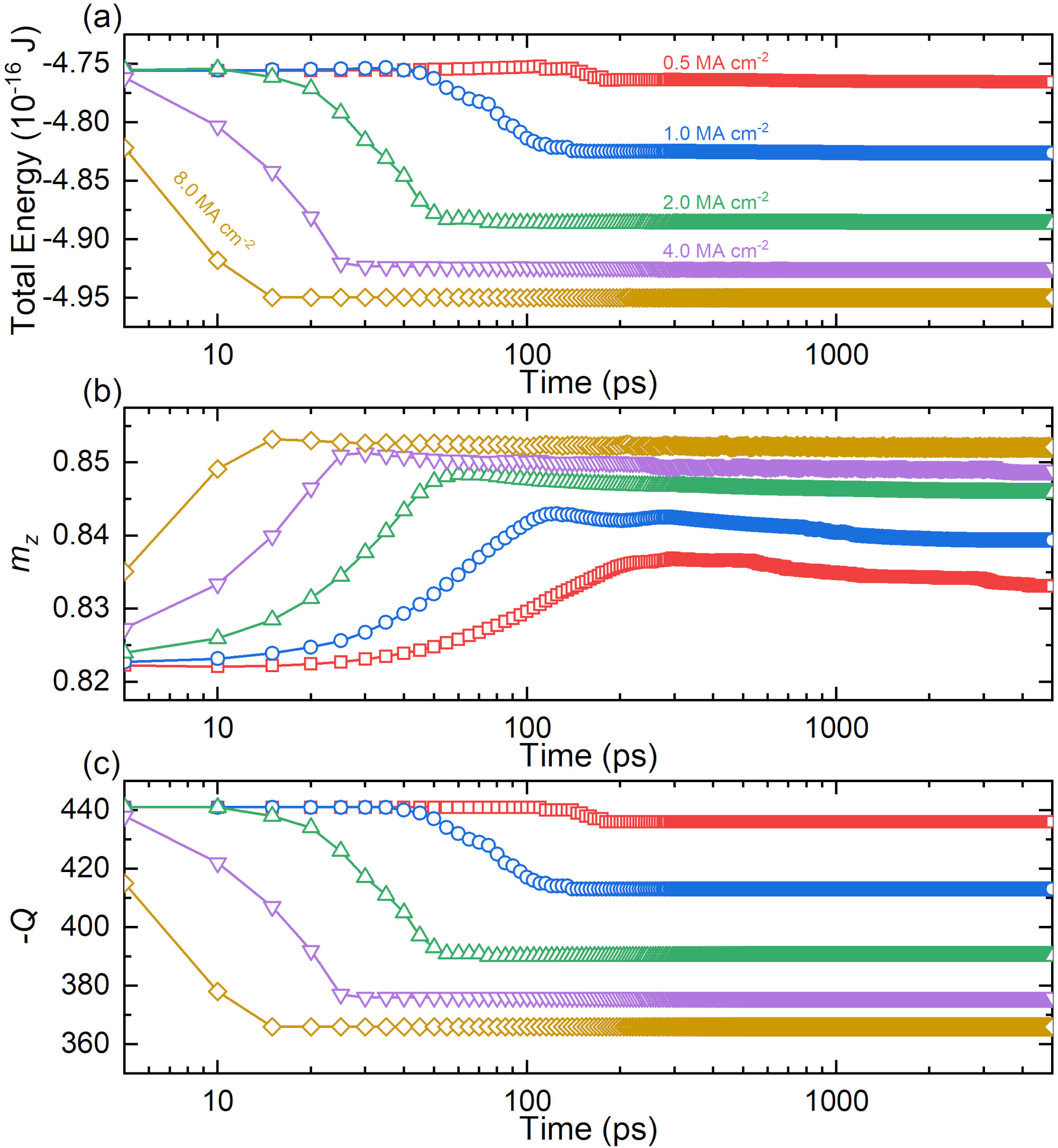}}
\caption{%
Time-dependent total energy, out-of-plane magnetization, and topological charge for the system with the \textit{initial state A} under compression.
(a) Time-dependent total energies for different driving current densities $j$.
(b) Time-dependent out-of-plane magnetization $m_z$ for different $j$.
(c) Time-dependent total topological charge $Q$ of the sample for different $j$.
Note that we show the additive inverse of $Q$. The current is turned on at $t=0$ ps in the driving region to compress skyrmions. The total simulation time is $4995$ ps.
}
\label{FIG2}
\end{figure}

\begin{figure}[t]
\centerline{\includegraphics[width=0.50\textwidth]{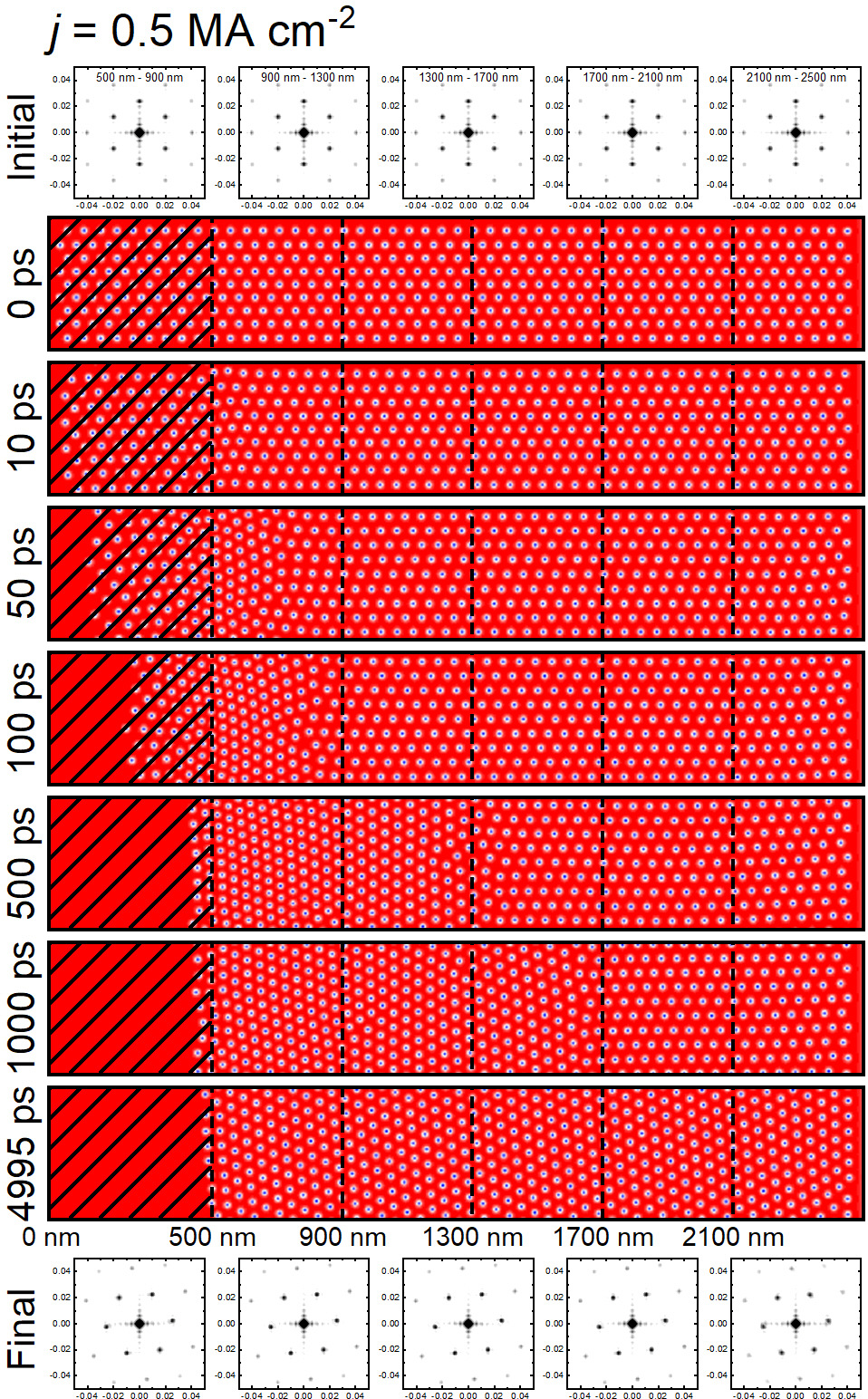}}
\caption{%
Selected snapshots and corresponding 2D Fourier transforms for the system with the \textit{initial state A} under compression.
A current of $j=0.5$ MA cm$^{-2}$ is applied in the driving region to compress skyrmions.
The current is turned on at $t=0$ ps, and the total simulation time is $4995$ ps. The 2D Fourier transforms of the spin configurations at $x=500-900$ nm, $x=900-1300$ nm, $x=1300-1700$ nm, $x=1700-2100$ nm, and $x=2100-2500$ nm are given for the initial ($t=0$ ps) and final ($t=4995$ ps) states, respectively.
}
\label{FIG3}
\end{figure}

\begin{figure}[t]
\centerline{\includegraphics[width=0.50\textwidth]{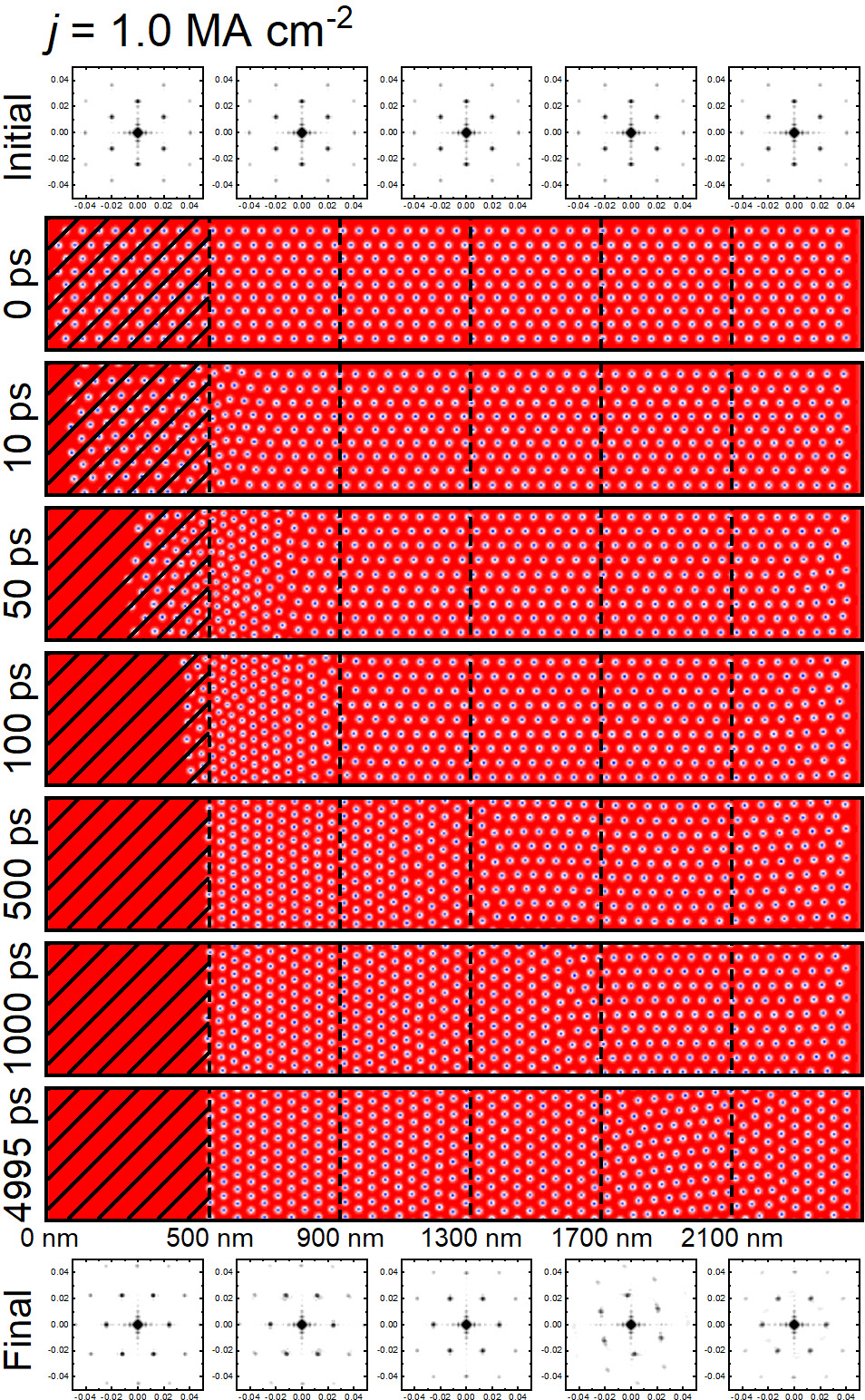}}
\caption{%
Selected snapshots and corresponding 2D Fourier transforms for the system with the \textit{initial state A} under compression.
A current of $j=1.0$ MA cm$^{-2}$ is applied in the driving region to compress skyrmions.
The current is turned on at $t=0$ ps, and the total simulation time is $4995$ ps. The 2D Fourier transforms of the spin configurations at $x=500-900$ nm, $x=900-1300$ nm, $x=1300-1700$ nm, $x=1700-2100$ nm, and $x=2100-2500$ nm are given for the initial ($t=0$ ps) and final ($t=4995$ ps) states, respectively.
}
\label{FIG4}
\end{figure}

\begin{figure}[t]
\centerline{\includegraphics[width=0.50\textwidth]{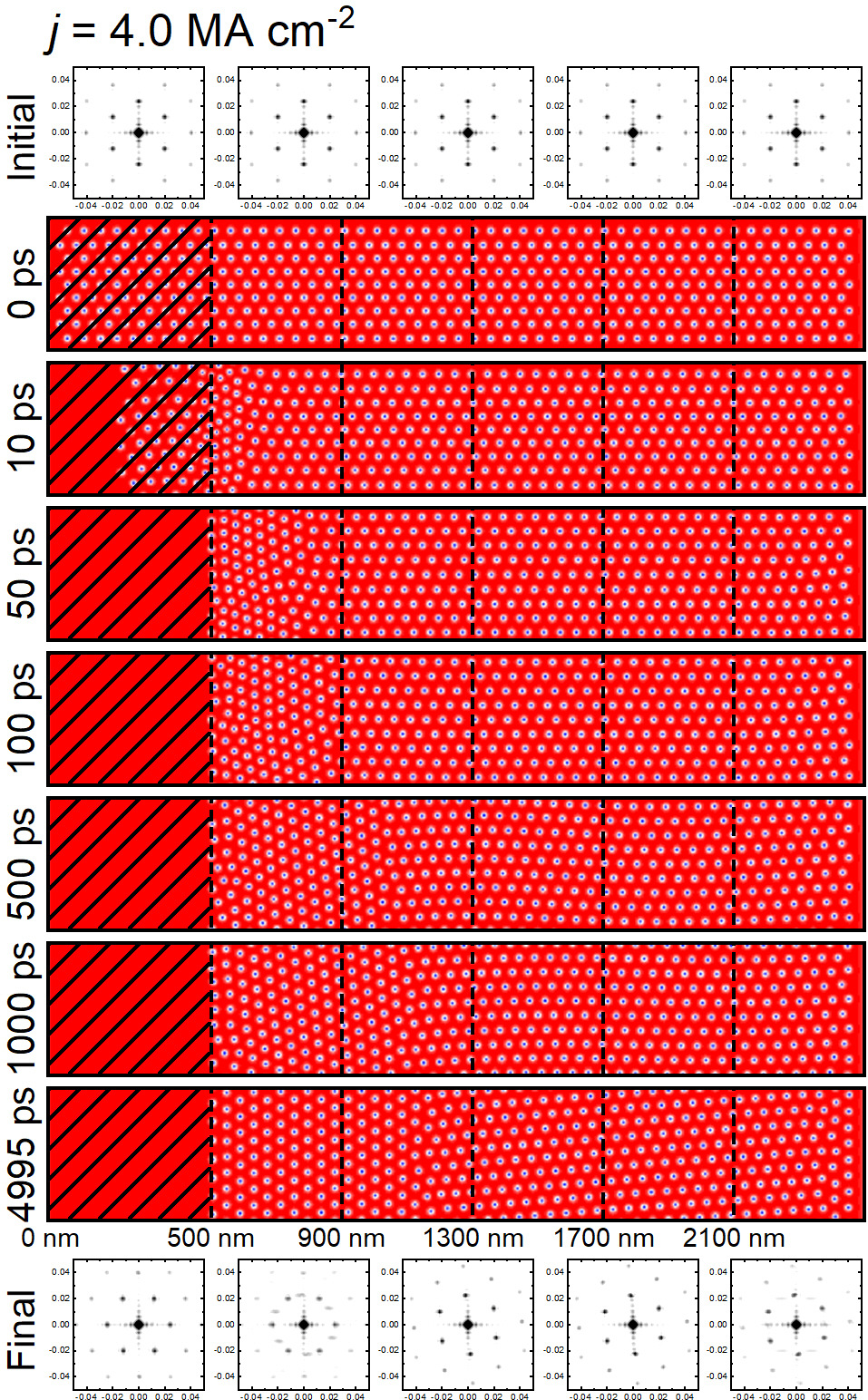}}
\caption{%
Selected snapshots and corresponding 2D Fourier transforms for the system with the \textit{initial state A} under compression.
A current of $j=4.0$ MA cm$^{-2}$ is applied in the driving region to compress skyrmions.
The current is turned on at $t=0$ ps, and the total simulation time is $4995$ ps. The 2D Fourier transforms of the spin configurations at $x=500-900$ nm, $x=900-1300$ nm, $x=1300-1700$ nm, $x=1700-2100$ nm, and $x=2100-2500$ nm are given for the initial ($t=0$ ps) and final ($t=4995$ ps) states, respectively.
}
\label{FIG5}
\end{figure}

\begin{figure}[t]
\centerline{\includegraphics[width=0.50\textwidth]{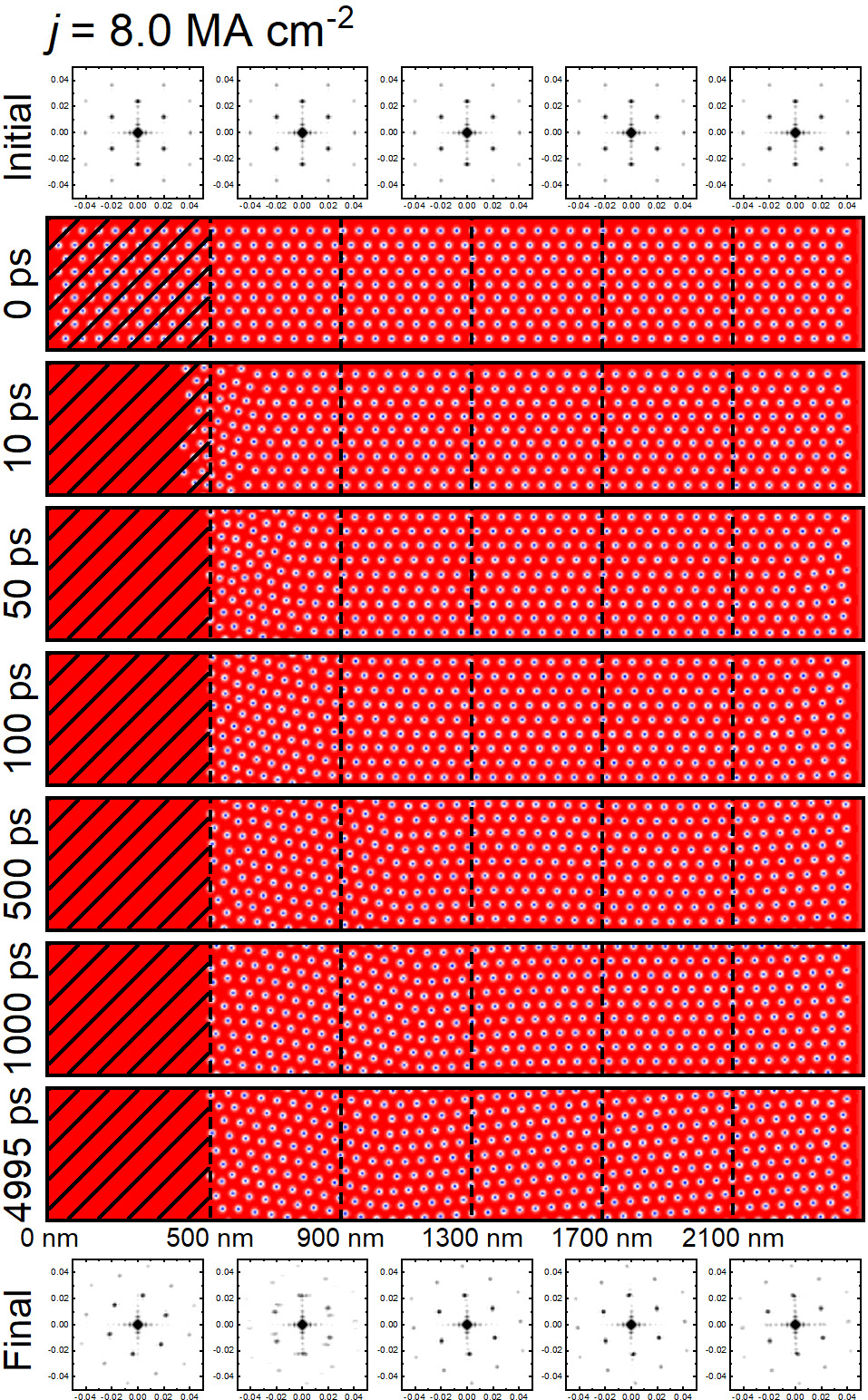}}
\caption{%
Selected snapshots and corresponding 2D Fourier transforms for the system with the \textit{initial state A} under compression.
A current of $j=8.0$ MA cm$^{-2}$ is applied in the driving region to compress skyrmions.
The current is turned on at $t=0$ ps, and the total simulation time is $4995$ ps. The 2D Fourier transforms of the spin configurations at $x=500-900$ nm, $x=900-1300$ nm, $x=1300-1700$ nm, $x=1700-2100$ nm, and $x=2100-2500$ nm are given for the initial ($t=0$ ps) and final ($t=4995$ ps) states, respectively.
}
\label{FIG6}
\end{figure}

\section{Results and Discussion}
\label{se:Results}

We first study the skyrmion dynamics under compression in the system with the \textit{initial state A}.
We focus on the skyrmion dynamics in the compressive region ($x=500-2500$ nm) when a current is applied to drive the skyrmions in the driving region toward the compressive region [Fig.~\ref{FIG1}(a)].
When the current in the driving region ($x=0-500$ nm) is turned on, all skyrmions in the driving region will move toward the compressive region (see Videos \blue{1}-\blue{4} in the Supplemental Material~\cite{SI}).
Due to the skyrmion Hall effect~\cite{Zang_PRL2011,Wanjun_NPHYS2017,Litzius_NPHYS2017,Reichhardt_NJP2016}, the skyrmions driven by the current have a positive velocity component in the length direction of the nanotrack (i.e., $v_x>0$) and a negative velocity component in the width direction of the nanotrack (i.e., $v_y<0$).
Therefore, when the skyrmions in the driving region meet and interact with the steady skyrmions in the compressive region, the collision between skyrmions around the skyrmion injection boundary between the driving and compressive regions (i.e., $x=500$ nm) first results in the compression of skyrmions near the injection boundary.
At the same time, as long as the current is turned on in the driving region, a chain of skyrmions remained in the driving region but near the injection boundary at $x=500$ nm shows a steady motion along the surface of the compressed skyrmions (see Videos \blue{1}-\blue{4} in the Supplemental Material~\cite{SI}).

In Fig.~\ref{FIG2}(a), we show the total micromagnetic energy of the system when the driving current is applied for $4995$ ps from $t=0$ ps.
As the compression effect is generally determined by the skyrmion speed in the driving region, which is controlled by the applied current density $j$, we apply a current of $j=0.5-8.0$ MA cm$^{-2}$ in the driving region to obtain different compression level.
We find that a larger current density could result in a faster skyrmion motion in the driving region and thus a faster and stronger collision between skyrmions near the injection boundary between the driving and compressive regions, however, a faster and stronger collision between skyrmions could lead to the annihilation of more skyrmions in the system.
Hence, the total micromagnetic energy decreases more significantly during the compression for the system driven by a larger current density.

The annihilation of skyrmions during the compression also means that the total number of skyrmions is reduced in the nanotrack, which will result in a relatively lower total skyrmion density in the compressive region, and thus, may lead to a weaker compression effect.
This also means the compressive effect is not simply proportional to the driving force for the skyrmion quasiparticle system.
Namely, the annihilation of skyrmion quasiparticles under compression creates a different compression phenomenon compared to the system of unbreakable particles, where the compression stress is proportional to the driving force.
The compression-induced annihilation of skyrmion quasiparticles could also be treated as a kind of structural failure of the skyrmion solid.
Therefore, in our system where no new skyrmions are added to the driving region during the compression, a moderate driving current may lead to a better compression level in the compressive region.

In Figs.~\ref{FIG2}(b) and~\ref{FIG2}(c), we show the out-of-plane magnetization component $m_z$ and total topological charge $Q$ of the nanotrack.
The compression-induced annihilation of skyrmions can also be seen from the time-dependent $m_z$ and $-Q$, which is more remarkable for a larger $j$.
However, it should be noted that when the system under compression reaches a stable number of skyrmions or total energy, it does not mean the system is in an equilibrium state.
The reason is that a density change of skyrmion quasiparticles induced by the compression may still be propagating along the nanotrack, which we will discuss by imaging the real-space spin configurations and analyzing the local skyrmion density later.

\begin{figure*}[t]
\centerline{\includegraphics[width=1.00\textwidth]{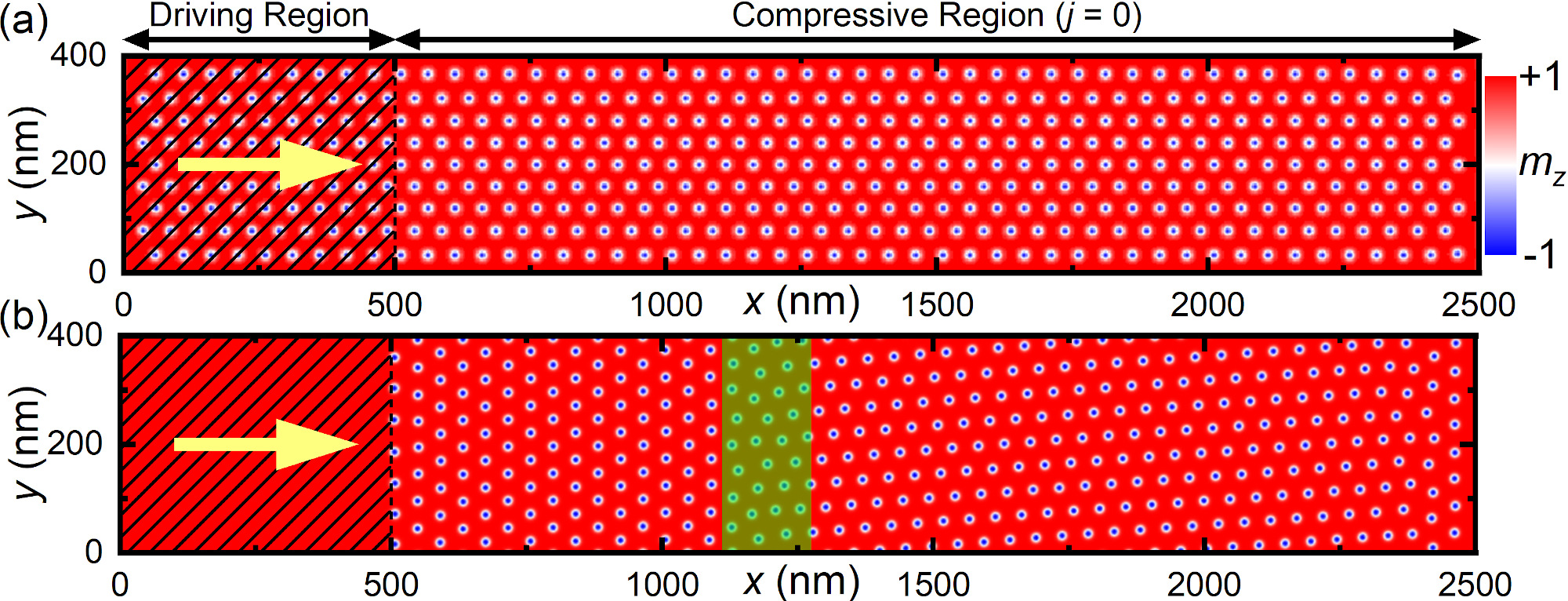}}
\caption{%
(a) Enlarged top view of the initial spin configuration at $t=0$ ps, corresponding to Fig.~\ref{FIG5}. The color scale represents the out-of-plane spin component $m_z$. A current of $j=4.0$ MA cm$^{-2}$ is applied in the driving region to compress skyrmions.
(b) Enlarged top view of the final spin configuration at $t=4995$ ps, corresponding to Fig.~\ref{FIG5}. The semitransparent green box highlights the transition region between different skyrmion lattice structures.
}
\label{FIG7}
\end{figure*}

In Figs.~\ref{FIG3}-\ref{FIG6}, we show the top-view snapshots of the nanotrack with the \textit{initial state A} under compression by different driving current densities.
For the case driven by $j=0.5$ MA cm$^{-2}$ [Fig.~\ref{FIG3}], only a few skyrmions are annihilated during the compression, which means that most skyrmions are pushed into the compressive region.
Such a compression first results in the interskyrmion interaction near the injection boundary at $x=500$ nm, which leads to an obvious increase of local skyrmion density as well as a structural transition of the skyrmion lattice.
For example, the skyrmion lattice structure at $x=500-1300$ nm is transformed from the \textit{initial state A} to a rotated triangular lattice at $t=500$ ps, which is similar to the \textit{initial state B} given in Fig.~\ref{FIG1}(b).

The structural transition of the triangular skyrmion lattice in the compressive region is caused by two reasons.
The first reason is that the rotated triangular lattice has a higher local skyrmion density favored by the compression.
The second reason is that the rotated triangular lattice has the shortest and almost identical spacing between adjacent skyrmions in the width direction of the nanotrack (i.e., the $y$ direction), which is the most stable and compact arrangement favored by the moving skyrmions near the injection boundary at $x=500$ nm (see Video \blue{1} in the Supplemental Material~\cite{SI}).

In Fig.~\ref{FIG3}, we can also see that an abrupt change of skyrmion density associated with the rotated triangular lattice structure propagates from the injection boundary toward the right end of the compressive region, which ultimately reaches the right end within $4995$ ps, leading to the full structural transition in the compressive region.
The different initial and final spin configurations of the skyrmion lattice in the compressive region can also be seen from the corresponding 2D Fourier transforms given in Fig.~\ref{FIG3}, where the pattern of six equally distributed first-order peaks is rotated almost $90$ degrees.
This also means that the triangular skyrmion lattice of the \textit{initial state B} is a spin configuration favored by the system under compression, which we will discuss later.

In Fig.~\ref{FIG4}, we show the case driven by a larger $j=1.0$ MA cm$^{-2}$.
As more skyrmions are annihilated under compression due to the skyrmion-skyrmion collision near the injection boundary at $x=500$ nm [Fig.~\ref{FIG2}(c)], the compression-induced structural transition effect does not penetrate into the full compressive region within $4995$ ps before the system reaches a dynamic equilibrium (see Video \blue{2} in the Supplemental Material~\cite{SI}).
Note that the dynamic equilibrium means that the structure of the compressed skyrmion solid remains unchanged while a chain of skyrmions may still move along the left surface of the skyrmion solid at $x\sim 500$ nm when the current is turned on in the driving region.
However, it is noteworthy that a structural transition of the skyrmion lattice happens near the right end of the compressive region ($x=2100-2500$ nm) at $t\sim 4995$ ps, although the structural transition effect does not reach the right end.
This phenomenon can be explained by the interaction between the right nanotrack edge and the skyrmion quasiparticles with a slight drift velocity.

\begin{figure*}[t]
\centerline{\includegraphics[width=0.85\textwidth]{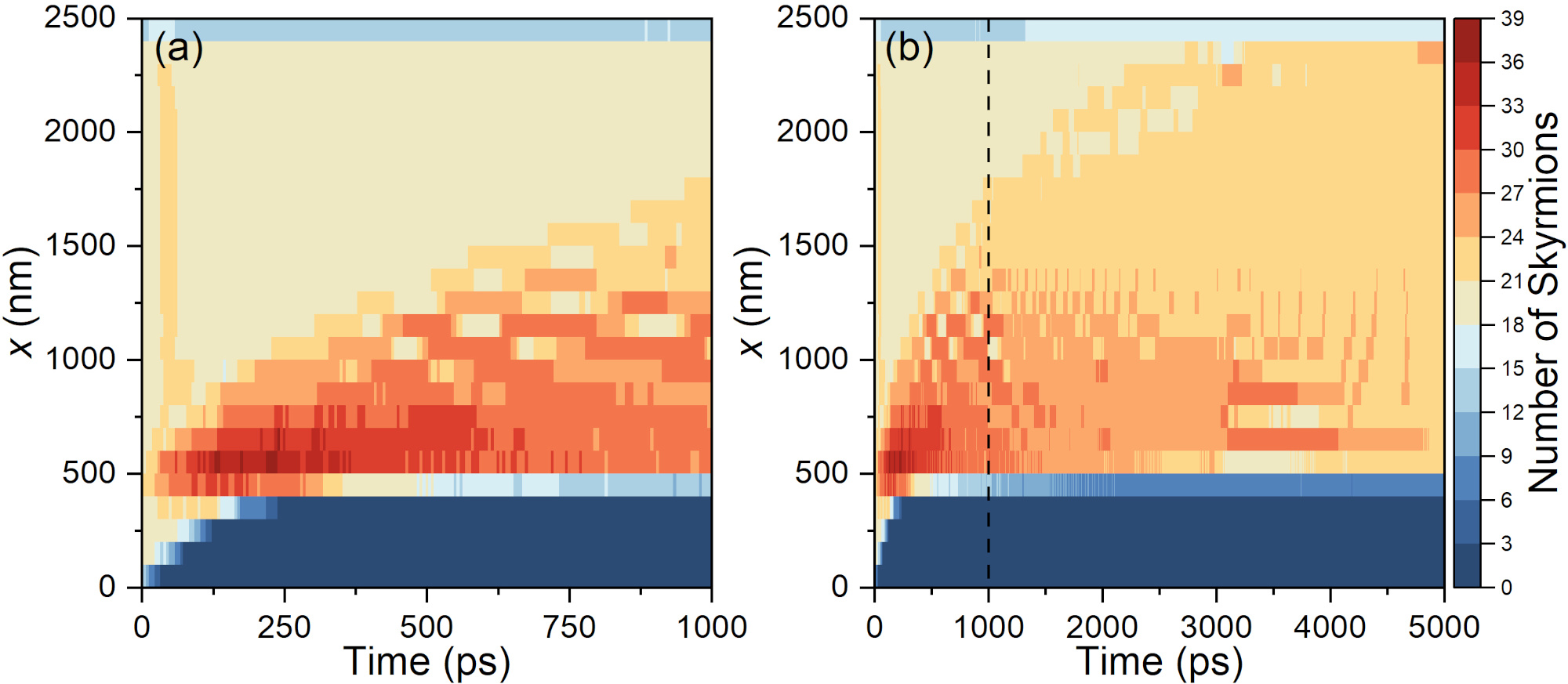}}
\caption{%
Spatiotemporal evolution of the skyrmion density for the system with the \textit{initial state A} under compression.
(a) Number of skyrmions in the nanotrack as functions of time ($t=0-1000$ ps) and the $x$ coordinate ($x=0-2500$ nm).
A current of $j=0.5$ MA cm$^{-2}$ is applied in the driving region to compress skyrmions. The current is turned on at $t=0$ ps, and the total simulation time is $4995$ ps.
The number of skyrmions is calculated every $100$ nm along the length direction (i.e., the $+x$ direction) of the nanotrack.
A larger local number of skyrmions per $100$ nm along the length direction indicates a higher local skyrmion density.
(b) Zoomed-out view of (a) for $t=0-4995$ ps.
The color scales are the same for panels (a) and (b).
}
\label{FIG8}
\end{figure*}

As demonstrated in Video \blue{2} in the Supplemental Material~\cite{SI}, a chain of skyrmions moves in the $-y$ direction along the left surface of the skyrmion solid under compression, which leads to a drift motion of the skyrmion solid in the compressive region in the $+y$ direction as an opposite reaction.
Such a drift velocity of the skyrmion quasiparticles along with a certain compression effect due to the overall increase of skyrmion density in the compressive region finally results in the mixed skyrmion lattice structure in the nanotrack with two separated areas being transformed.

In Fig.~\ref{FIG5}, we show the case driven by a much larger $j=4.0$ MA cm$^{-2}$.
Initially, we note that some non-six fold coordinated skyrmions are formed near the injection boundary at $t=10$ ps, where the skyrmion density is changing most rapidly.
Many skyrmions in the driving region are annihilated within the first $100$ ps due to the strong collision driven by the current.
The compression level is much reduced due to the reduced total number of skyrmions that can be compressed in the compressive region.
It can be seen that the compression-induced structural transition of the skyrmion lattice structure is realized only for $x<1300$ nm when the system reaches the dynamic equilibrium before $t\sim 4995$ ps (see Video \blue{3} in the Supplemental Material~\cite{SI}).

In Fig.~\ref{FIG6}, we show the case driven by a strong current of $j=8.0$ MA cm$^{-2}$.
The strong current drives the skyrmions in the driving region into very fast motion toward the compressive region, and results in destructive collisions among skyrmions around the injection boundary.
Therefore, most skyrmions in the driving region are annihilated within the first $50$ ps before being pushed into the compressive region.
In such a case, the skyrmion quasiparticles in the compressive region are slightly compressed and almost affected by a chain of skyrmions moving in the $-y$ direction along the left surface of the compressed skyrmion quasiparticles (see Video \blue{4} in the Supplemental Material~\cite{SI}). As a result, a slight distortion of the skyrmion lattice structure in the compressive region is achieved.
In Fig.~\ref{FIG7}, we show the enlarged view of the initial and final spin configurations for the case driven by $j=4.0$ MA cm$^{-2}$ [Fig.~\ref{FIG5}], where we highlight the transition region between two final skyrmion lattices with different orientations. It can be seen that the structure of the skyrmion lattice within the transition region is obviously distorted.

We further analyze the spatiotemporal evolution of the local skyrmion density for the system with the \textit{initial state A} under compression by a small current of $j=0.5$ MA cm$^{-2}$.
We focus on the case driven by the small current $j=0.5$ MA cm$^{-2}$ as the compression-induced structural transition can penetrate into the whole compressive region [Fig.~\ref{FIG3}].
It can been seen from Fig.~\ref{FIG8} that the applied current drives most skyrmions in the driving region (i.e., $x=0-500$ nm) into the compressive region (i.e., $x=500-2500$ nm) within the first $250$ ps, leading to an abrupt increase of the local skyrmion density near the skyrmion injection boundary at $x=500$ nm.
The local skyrmion density near the injection boundary then decreases and a wave of compression is formed, which is indicated by the increase of the local skyrmion density from the injection boundary toward the area near the right end of the compressive region.
Namely, the wave front of the skyrmion compression propagates along the length direction of the nanotrack.
It should be noted that the propagation of wave front may be affected by the skyrmion Hall effect, so that during compression the front motion would not be strictly in the length direction. Such an effect could be examined by assuming a much smaller damping parameter.

In particular, we also note that the abrupt increase of skyrmion density in the narrow area near the injection boundary (i.e., $x=500-750$ nm) followed by a slow decrease of skyrmion density may be understand as a kind of compression wave.
The spike of such a wave in the compressed skyrmion quasiparticles does not travel well and is damped out by the large damping effect of the ferromagnetic nanotrack.
However, we do not observe a rarefaction wave in our system.
One main reason is that the total number of skyrmions in the compressive region is increasing and approaches a higher number when the system is under compression, which means that the overall density of skyrmions will be increased in the compressive region so that the skyrmion dynamics is dominated by strong skyrmion-skyrmion repulsive interactions.
In such a system with dense skyrmions and strong interskyrmion repulsion, the reduction of local skyrmion density compared to the initial state in the compressive region cannot be produced.
Another possible reason for the absence of rarefaction wave in our system is that the system has no or tiny inertia.

In this work, we also study the compression dynamics of skyrmion quasiparticles in the nanotrack with the \textit{initial state B}, which is a triangular skyrmion lattice configuration favored by the compression effect, as demonstrated in Figs.~\ref{FIG3}-\ref{FIG6}.
We find that the time-dependent behaviors of the total energy, out-of-plane magnetization component, and total topological charge during the compression are almost the same as those of the system with the \textit{initial state A} (see Supplemental Fig.~\blue{1} in the Supplemental Material~\cite{SI}).
However, it can be seen from the top-view snapshots of the nanotrack that the compression effect does not induce a structural transition of the given triangular skyrmion lattice (see Supplemental Fig.~\blue{2} and Videos \blue{5}-\blue{8} in the Supplemental Material~\cite{SI}).
Indeed, the compression effect could result in the increase of skyrmion density in the compressive region, and may also cause certain level of lattice distortion (see Supplemental Fig.~\blue{3} in the Supplemental Material~\cite{SI}).

In the Supplemental Material~\cite{SI}, we also show the results of the nanotrack with the \textit{initial state C} under compression by different driving current densities.
For such a system with a square skyrmion lattice, the time-dependent variations of the total energy, out-of-plane magnetization component, and total topological charge during the compression are qualitatively similar to those of the system with a triangular skyrmion lattice as the initial state.
However, we find that the square skyrmion lattice in the whole compressive region will transform to a triangular skyrmion lattice similar to the \textit{initial state A} soon upon the application of the compression from the driving region (i.e., within about the first $50$ ps), as shown in Supplemental Fig.~\blue{5} in the Supplemental Material~\cite{SI}; also see Videos \blue{9}-\blue{12} in the Supplemental Material~\cite{SI}.

Such a fast transition of the skyrmion lattice structure in the whole compressive region at the initial stage of compression is a result of the skyrmion-skyrmion repulsions between neighboring skyrmions as well as the initial compression when some skyrmions in the driving region are pushed into the compressive region through the injection boundary at $x=500$ nm.
The initial rearrangement of the skyrmion lattice structure in the whole compressive region favors a more stable triangular lattice with compact spacing between adjacent skyrmions in the length direction of the nanotrack (i.e., the $x$ direction), which justifies that the skyrmion-skyrmion repulsion is enhanced in the $x$ direction due to the general compressive force normal to the injection boundary at $x=500$ nm.
When more skyrmions are pushed into the compressive region, a structural transition of the initially formed triangular skyrmion lattice is produced, forming a triangular lattice similar to the \textit{initial state B}.

It is worth mentioning that the structural transition between different triangular skyrmion lattices are mainly caused by the slow drift motion of skyrmion quasiparticles under compression as discussed above, while the square-to-triangular structural transition is caused by the skyrmion-skyrmion repulsion in the system under compression.
We emphasize again that the drift motion of the skyrmion quasiparticles in the compressive region is induced by the motion of a chain of skyrmions near the injection boundary, where the skyrmions remained in the driving region are driven by the current and guided along the left surface of the skyrmion quasiparticles in the compressive region.
We also note that the drift motion of the skyrmion quasiparticles in the compressive region may result in multiple structural transitions (see Video \blue{9} in the Supplemental Material~\cite{SI}), which is a phenomenon depending on the arrangement of skyrmions on the compressed skyrmion solid surface where the motion of skyrmion chain is guided.

\section{Conclusion}
\label{se:Conclusion}

In conclusion, we have studied the skyrmion dynamics in a nanotrack under compression.
The compression is realized by pushing skyrmions from the left driving region into the right compressive region using the dampinglike spin torque, which could lead to an abrupt increase of skyrmion density in the compressive region and also a motion of a chain of skyrmions near the injection boundary separating the driving and compressive regions.
The compression may also lead to the annihilation of some skyrmions around the injection boundary due to the strong collision between skyrmions.

We find that the compression effect could result in a structural transition of the skyrmion lattice in the compressive region.
The structural transition depends on both the compression level as well as the initial skyrmion lattice structure in the compressive region.
In particular, we find that the compression-induced transition between two different triangular skyrmion lattice structures is subject to both the change of skyrmion density and the drift motion of skyrmions, while the compression-induced square-to-triangular lattice transition is mainly subject to the skyrmion-skyrmion repulsion in the direction normal to the injection boundary.

The transition between two different triangular skyrmion lattice structures propagates along the direction normal to the skyrmion injection boundary and could reach the right end of the nanotrack if the compression level is strong enough.
In contrast, the square-to-triangular lattice transition in the compressive region happens in a collective manner and a much faster time scale.

At last, we point out that the compression front structures formed during the compression in our system may be affected by the sample size in the $y$ dimension, although periodic boundary conditions are applied in the $y$ direction. The possible propagation of compression front in the $y$ direction due to the skyrmion Hall effect may not be observed when the sample has a fairly small finite size in the $y$ direction.
Our results could be useful for the understanding of skyrmion quasiparticle dynamics under compression and may open a new way for the manipulation of skyrmion lattice structure. 

\begin{acknowledgments}
X.Z. was an International Research Fellow of the Japan Society for the Promotion of Science (JSPS).
X.Z. was supported by JSPS KAKENHI (Grant No. JP20F20363).
J.X. was a JSPS International Research Fellow.
J.X. was supported by JSPS KAKENHI (Grant No. JP22F22061).
X.L. acknowledges support by the Grants-in-Aid for Scientific Research from JSPS KAKENHI (Grants No. JP20F20363, No. JP21H01364, No. JP21K18872, and No. 22F22061).
\end{acknowledgments}



\end{document}